\documentclass[namedreferences,hyperref,optionalrh]{spr-sola}
\usepackage{graphicx}        
\usepackage{amssymb}        
\usepackage{color}           
\usepackage{breakurl}                         

\chardef\us=`\_

\begin{document}
\begin{opening} 

\title{Meridional Movements of Individual Sunspots and Pores}

\author[addressref=aff1,email={tlatov@mail.ru}]{\inits{A.G.}\fnm{~Andrey~G.}~\lnm{~Tlatov}\orcid{0000-0002-6286-3544}}
\author[addressref=aff1,email={}]
{\inits{K.A.}\fnm{~Kseniya~A.}~\lnm{Tlatova}\orcid{0000-0001-8345-6801}}

\address[id=aff1]{Kislovodsk Mountain Astronomical Station of the Pulkovo observatory, Kislovodsk, Gagarina str. 100, 357700, Russia.}

\runningtitle{Meridional Movements of Individual Sunspots and Pores}
\runningauthor{A.G. Tlatov and K.A. Tlatova}

\begin{abstract}
The analysis of the meridional displacement velocity of individual solar pores and sunspots has been performed. In the period May 2010\,-\,March 2025 of observations in the continuum of the  \textit{Solar Dynamics Observatory/Helioseismic and Magnetic Imager} (SDO/HMI), we identified more than $3.6\cdot 10^{5}$  sunspots and pores for analysis and tracked their displacement. The velocity of the meridional displacement of spots $v_{\rm m}$ depends on their magnetic polarity, latitude, and stage of development. For sunspots and pores of trailing polarity, the velocity of movement is on average directed toward the poles. For such spots, the dependence of the velocity on latitude can be represented by linear regressions for pores: $v^{\rm pr}_{\rm tr} \approx 2.0+0.62\cdot \theta^{\rm o}$  m s$^{-1}$; for sunspots: $v^{\rm sp}_{\rm tr}\approx 0.02+0.94\cdot \theta ^{\rm o}$ m s$^{-1}$. 
For sunspots and pores of leading polarity, the dependence is non-monotonic in nature on latitude. For pores: $v^{\rm pr}_{\rm ld}\approx 0.35-11.7\cdot {\rm sin}(\theta)+16.5\cdot {\rm sin}^{\rm 2} (\theta) +76.5\cdot {\rm sin}^3 (\theta)-32.7\cdot {\rm sin}^{4}(\theta )$ m s$^{-1}$;  for sunspots: $v^{\rm sp}_{\rm ld}\approx -0.35-18.3\cdot {\rm sin}(\theta)+32.2\cdot {\rm sin}^{\rm 2}(\theta) +71.4\cdot {\rm sin}^{\rm 3} (\theta)-6.7\cdot {\rm sin}^{\rm 4}(\theta)$  m s$^{-1}$. 
The highest speed of meridional movement to the poles is observed for sunspots of trailing polarity during the phase of growth of the sunspot area. The velocity of the meridional movement depends on their area, reaching a maximum for an area of $S\approx 80 - 100$ $\mu$sh.

\end{abstract}
\end{opening} 

\section{Introduction} \label{sec:intro}

There are different approaches to studying the meridional movement of sunspots. Some authors associate movement with global meridional circulation. The solar meridional circulation is an axisymmetric system of flows that extend from the equator to the poles at $v_{\rm m}\approx 20$ m s$^{-1}$ on the surface \citep{Hanasoge}. The most widely used data for analysis are those on sunspot groups. Analysis of data from the Royal Greenwich Observatory (RGO) in the period 1874\,--\,1935 \citep{Tuominen} showed that the motion of sunspot groups is directed toward the equator at low latitudes and toward the pole at high latitudes, reaching values of 0.1 deg/day or 14 m s$^{-1}$. \cite{Ward} using RGO data found velocities of $v_{\rm m}\approx 1$ m s$^{-1}$. Other studies came to different conclusions; for example, \cite{Ribes} found large alternating flows of the order of 100 m s$^{-1}$. \cite{Howard86} found velocities in the polar and equatorward directions. \cite{Kambry} measured velocities of 20 m s$^{-1}$ in the equatorial direction. In \cite{Arevalo} the authors used data on the position of sunspot groups for the period 1874\,--\,1902 from the RGO and found that in the equatorial region the spots move toward the poles and at latitudes above $(\theta \approx 10^{\rm o})$ toward the equator. In \cite{Coffey}, it was found that spots near the equator $(\theta <10^{\rm o})$ move mainly toward the equator.
\cite{Javaraiah} investigated the velocity of meridional sunspot groups as a function of the group lifetime and the phase of the solar cycle. The meridional velocities of sunspot groups aged 10\,-\,12 days are directed towards  to the poles  $\approx 10\,-\,20$ m $s^{-1}$. In \cite{Sivaraman}, the velocities of meridional movement were considered depending on the area of  sunspot groups based on observations from the Kodaikanal and Mount Wilson observatories. The highest velocities of meridional displacements toward the poles were found for groups of relatively small area 0\,-\,5 $\mu$sh ($\mu$sh - area in millionths of the solar hemisphere). 
The authors linked the changes in velocity with the area with different depths of the anchor of sunspot groups.

Another approach is that the meridional motions are characteristic of individual sunspots and may differ from the motion of the entire group. During the emergence of sunspot groups, the centers of their leading and trailing parts diverge \citep{Gilman86}. This divergent motion leads to the stretching of the active regions. \cite{Murakozy24} found that the greatest separation between spots of different polarity occurs during the area decay phase. \cite{Schunker} studied the motion of regions of different polarity and pointed out that two phases of sunspot emergence can be distinguished. During the first phase, the polarity separation accelerates, while during the second phase it slows down. One feature is that the leading part moves equatorward and westward, while the trailing part moves poleward and eastward. \cite{McClintock} studied the changes between the separation distances in sunspot groups of different sizes during their evolution, finding that the expansion behavior was different in sunspot groups of different sizes. In \cite{TT24} it was found that the rotation speed of sunspots and pores of leading and trailing polarity differs and it was suggested that this is due to additional movement of sunspots of different polarity during the rise of the magnetic field tube.  

In this study, we examine the meridional motion velocity of individual sunspots based on statistical analysis over a large sample.

\begin{figure}
\centerline{\includegraphics[width=1.0\textwidth,clip=]{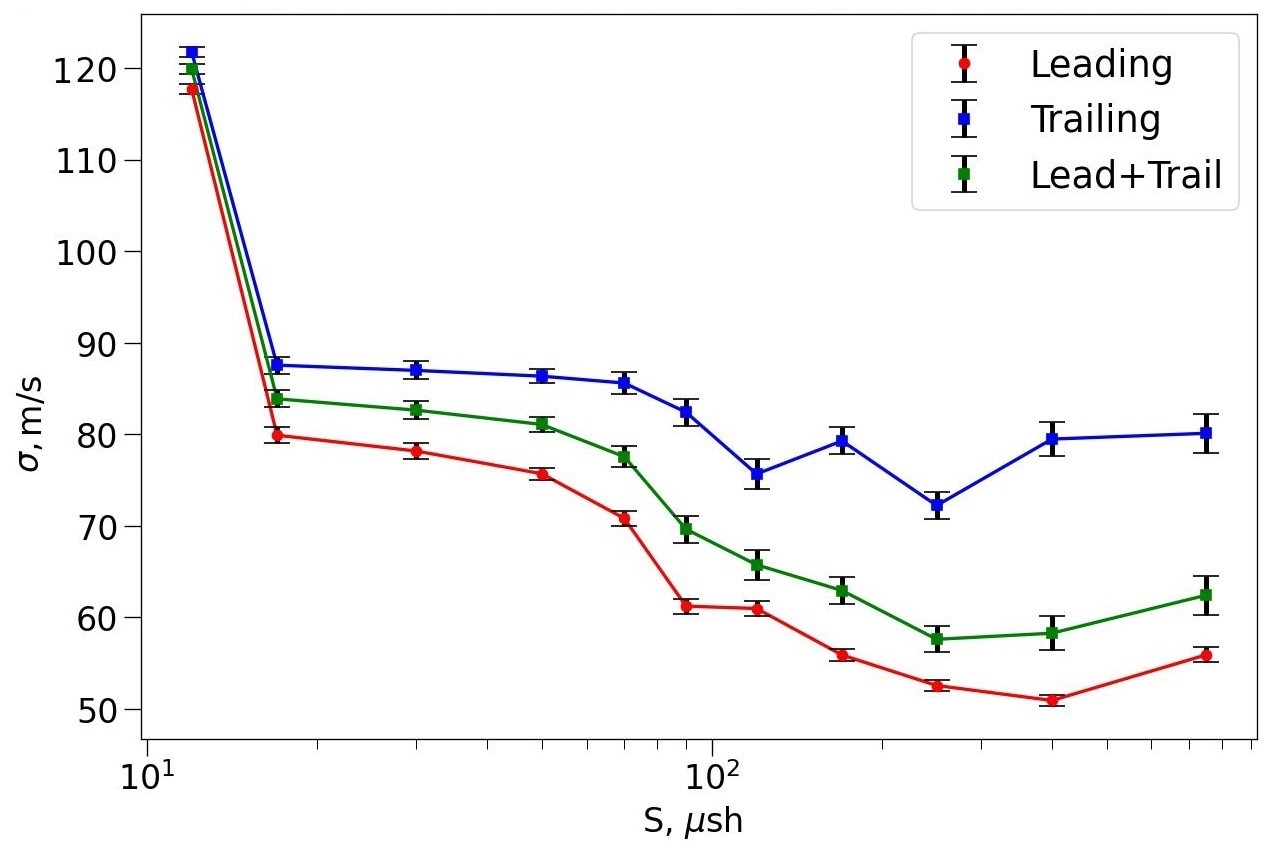}
                 } 
\caption{Standard deviation  of the meridional motion velocity measurements as a function of sunspot area. Data are presented for all spots ({\it green line}) and separately for spots of leading and trailing polarity. Confidence intervals are represented by {\it bars}.}
\label{fig1}
\end{figure}

\section{Data and Method}
For the analysis, we used the characteristics of individual sunspots and pores. For this, we analyzed the observations of \textit{Solar Dynamics Observatory/Helioseismic and Magnetic Imager} (SDO/HMI)  observatory of the \textsf{hmi.Ic\_45s} and \textsf{hmi.M\_45s} series, performed at the same time. We took 5 images for each day at times close to 00:00, 05:00, 10:00, 15:00 and 20:00 UT. The main data for processing were images in “white” light (\textsf{hmi.Ic\_45s}), in which sunspots and sunspot umbra were identified. The observation data were processed automatically. To identify sunspots and pores, we used a sunspot detection procedure similar to that used at the Kislovodsk Mountain Astronomical Station (KMAS) for identifying sunspots for synoptic observations \citep{Tlatov14, Tlatov19}.  

We identified sunspots and pores near the coordinates of the sunspot group centers using Kislovodsk data (\url{93.180.26.198:8000/web/Soln_Dann/}). The procedure for determining sunspot boundaries was described in \cite{TT24}. In this analysis, we used the characteristics of sunspots and pores in the period 1 May 2010\,--\,29 March 2025. More than 108 thousand sunspots and more than 250 thousand pores were identified.  Here by pores we mean small-area spots for which it is not possible to distinguish penumbra. As a rule, the area of solar pores is limited to $S<20$ $\mu$sh.

Using these data, we tracked the evolution of the spots using their coordinates in sequence. To form chains, we used the procedure of tracking sunspots identified in images adjacent in time. To search for spots, we selected spots with the same magnetic polarity, located at a minimum heliographic distance, and differing little in area. To determine the polarity of the magnetic field, we superimposed the boundaries of the spots on the observations of magnetic fields (\textsf{hmi.M\_45s}) and determined the maximum magnetic field along the line of sight. Pores in which the magnetic field intensity was $B<100$ G were excluded from consideration. 

To form chains, we used the procedure of tracking sunspots selected in neighboring-time images. To search for spots, we tracked spots that had the same magnetic polarity, were at a minimum heliographic distance, and differed little in area. To do this, we search for spots in neighboring images and had a minimum of the parameter: $d_{ij}=\sqrt{(\phi_{i}-\phi_{j})^{\rm 2}+(\theta_{i}-\theta_{j})^{\rm 2}}+(\Delta S)^{\rm 2}$, where $\theta,\phi$ are the latitude and longitude of the spots $i,j$. We corrected the longitude of spots on the subsequent image for differential rotation $\phi_{j}=\phi_{j}'-(13.198+\Delta t/24 + \Delta \omega)$,  where $\Delta t$  is the time between observations in hours, the term $\Delta \omega$ takes into account the differential rotation in latitude $\Delta \omega=0.44-3 \Delta t sin^{\rm 2}\theta$. The functional of the area was calculated as: $\Delta S =3 |S_{i}-S_{j}|/(0.5( S_{i}+S_{j})+20+2 \Delta t)$, here $S_{i}$ and $S_{j}$ are the area of spots in  $\mu$sh.  The spots were considered identical if their coordinates were less than $|\theta_{i}-\theta_{j}|<2+\Delta \theta/2$  and $|\phi_{i}-\phi_{j}|<2+\Delta \phi/2$, where $\Delta \theta$, $\Delta \phi$ are the extent of the spots in latitude and longitude, and the minimum of the parameter $d_{ij}$ was found for the pair ${ij}$. If spot $j$ could not be found for spot $i$ in the adjacent image, the sequence chain was interrupted. 
The velocity of the meridional displacement was found as $v_{\rm m}=(\theta_{i}-\theta_{j})/\Delta t$.

The data have a fairly large standard deviation $\sigma$, significantly larger than the expected rates. This limitation can be overcome by considering a significant number of measurements. Figure \ref{fig1} shows  $\sigma$ and the confidence interval of the mean $\sigma/\sqrt{n+1}$, where $n$ is the number of samples. The standard deviation value increases as we approach the limb. Therefore, in Figure \ref{fig1} and the subsequent analysis, we limited the distance from the disk center to $r<0.6R_\odot$. It can be seen that for small areas $S<15$ $\mu$sh, the dispersion increases significantly. This may be due to the fact that such an area is typical for solar pores, which are morphologically different from sunspots.

In total, we identified $\approx 330$ thousand pairs of spots in adjacent images and formed $\approx 72$ thousand chains to track spots over time.

\begin{figure}
\centerline{\includegraphics[width=1.0\textwidth,clip=]{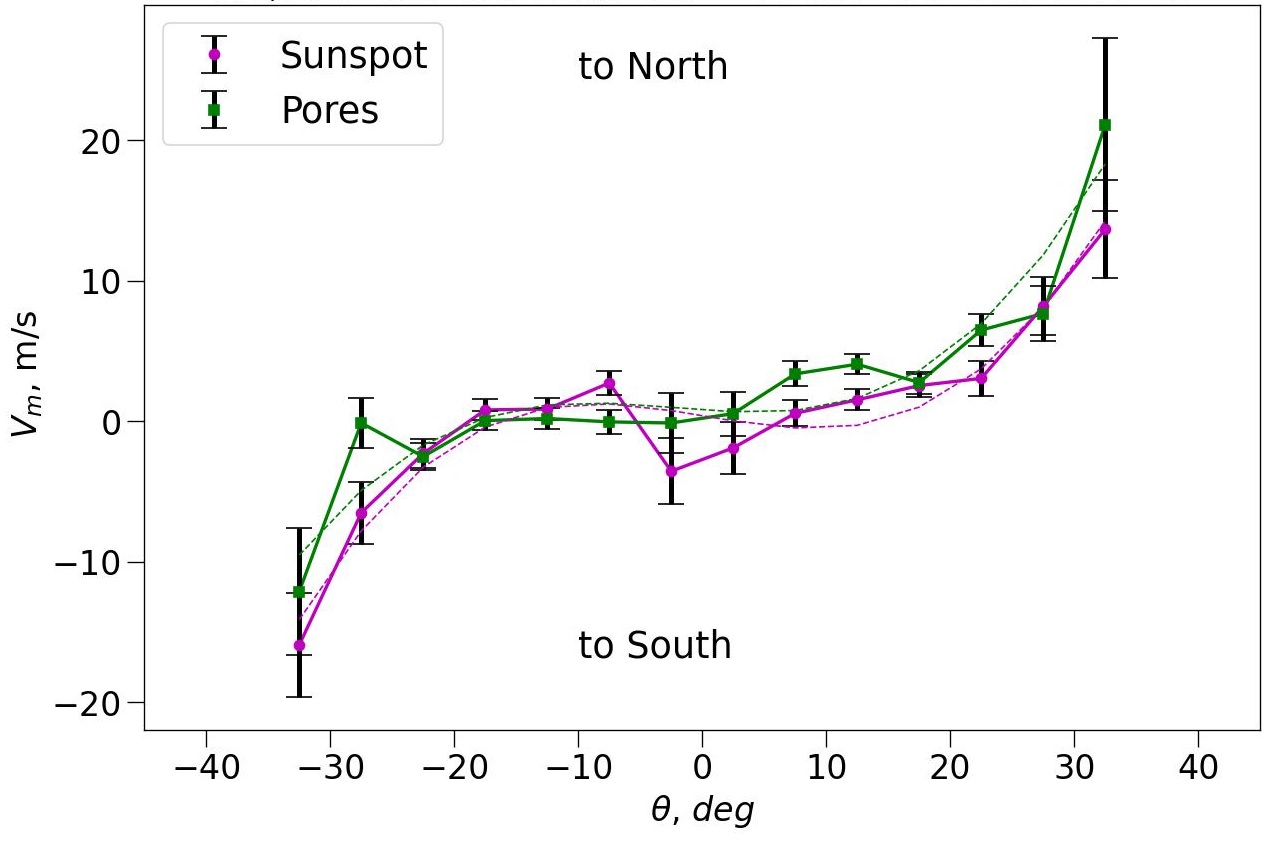}
                 } 
\caption{Meridional velocity of sunspots and pores.  The {\it dotted-line} represents the approximations. The {\it bars} show the confidence intervals.}
\label{fig2}
\end{figure}

\begin{figure}
\centerline{\includegraphics[width=1.0\textwidth,clip=]{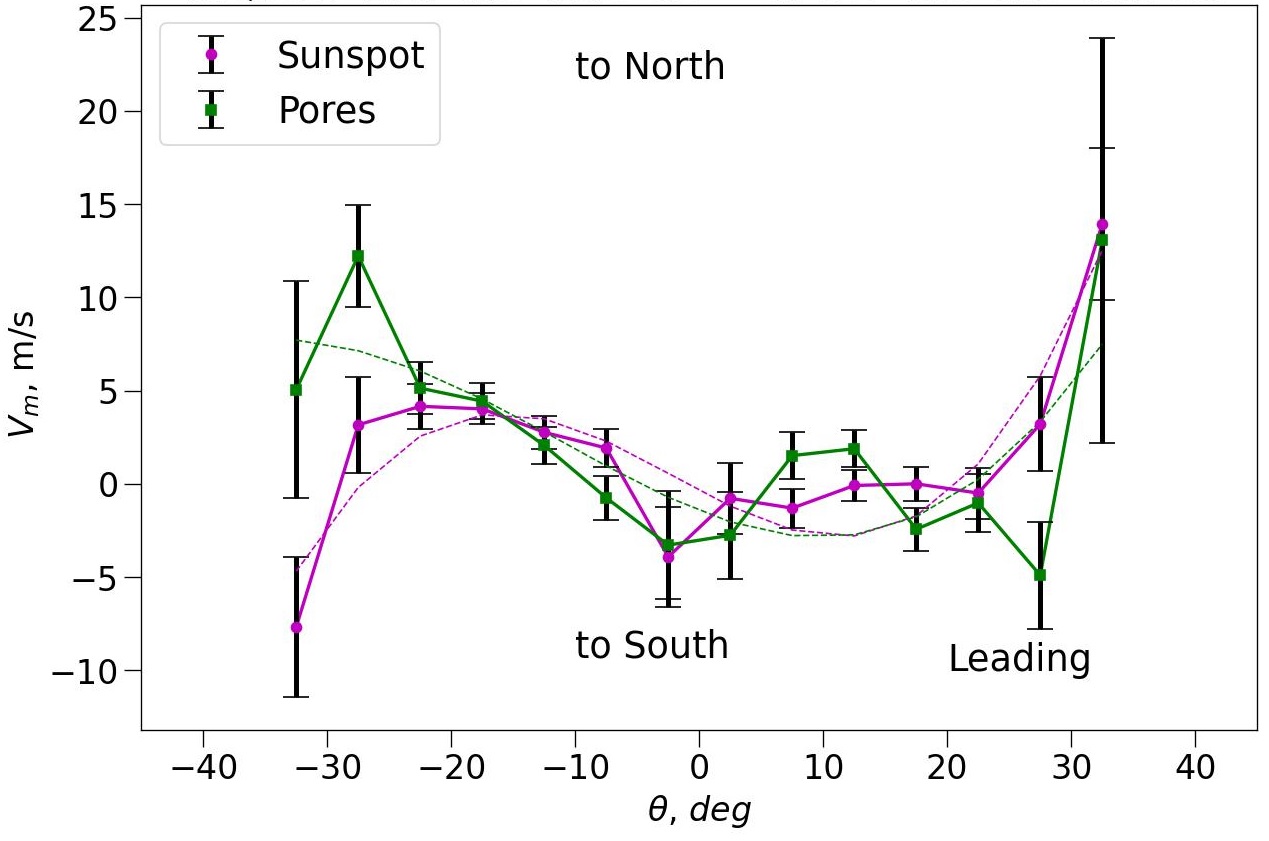}
                 } 
\caption{The same as Figure ~\ref{fig3}, but for spots of leading polarity.}
\label{fig3}
\end{figure}

\begin{figure}
\centerline{\includegraphics[width=1.0\textwidth,clip=]{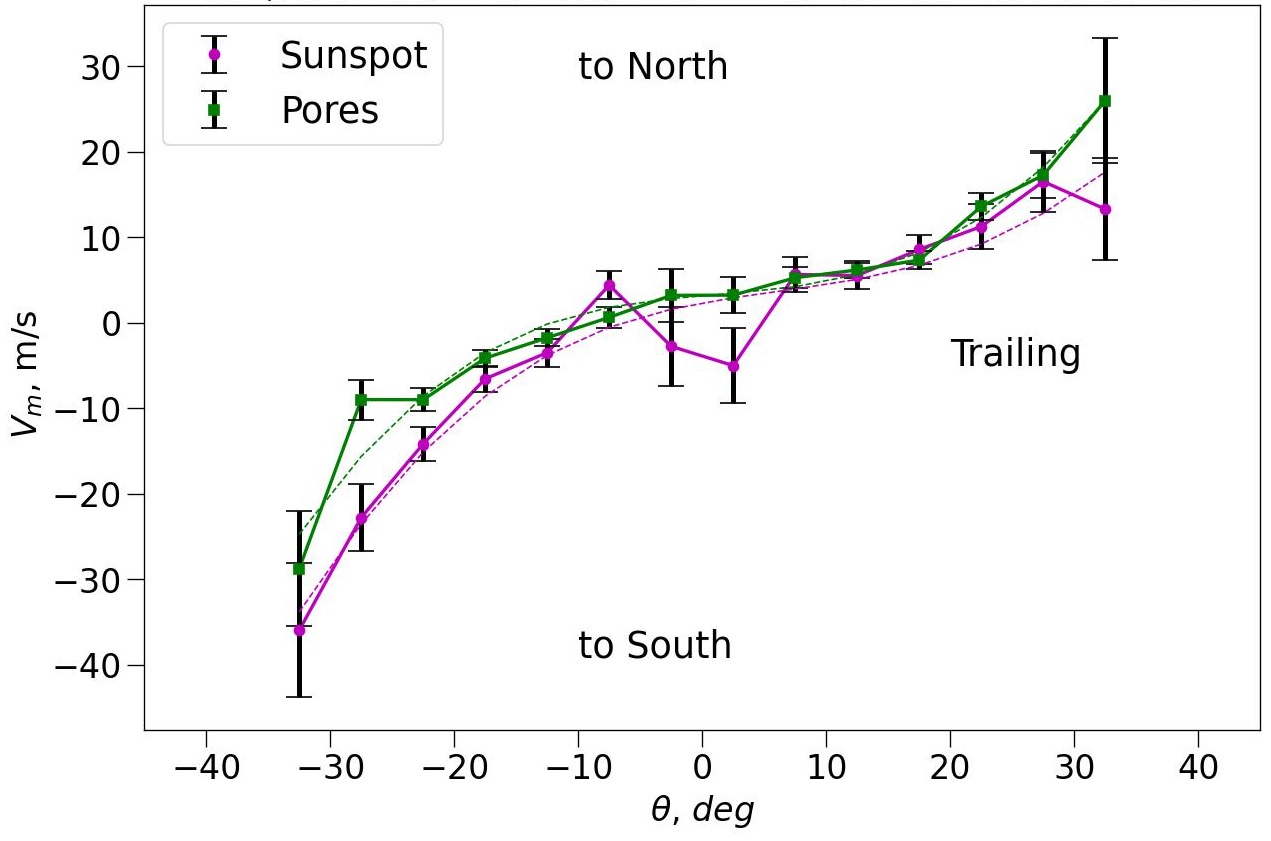}
                 } 
\caption{The same as Figure ~\ref{fig1}, but for spots of trailing polarity.}
\label{fig4}
\end{figure}

\section{Analysis Results}
Figure \ref{fig2} shows the dependence of the average meridional velocities of pores and sunspots on the heliographic latitude. To separate these two types of spots, we took sunspots with an area of at least $S>30\ \mu$sh. For the sample we limited the amplitude of meridional velocities $|v_{\rm m}|<300$ m s$^{-1}$ $\approx 3\sigma$. Figure \ref{fig2} shows all spots, without taking into account the magnetic polarity. The dependences of the velocities on latitude for sunspots and pores are close to each other. Near the equator, the velocities are close to zero. At latitudes $\theta >20^{\rm o}$  in the northern and southern hemispheres, the velocity is directed toward the poles, reaching a value of $v_{\rm m} \approx 10 - 15$  m s$^{-1}$.
The dependence of the speed of meridional movements on latitude can be represented as a 3rd degree polynomial of the sine of latitude $\theta$ for pores:  $v^{\rm pr}\approx~ 0.76-5.31\cdot{\rm sin}(\theta)-3.21\cdot \rm{sin}^{\rm 2} (\theta) +86\cdot \rm{sin}^{\rm 3} (\theta)$ m s$^{-1}$, for sunspots: $v^{\rm sp}\approx 0.84-5.08\cdot \rm{sin}(\theta)-0.69\cdot {\rm sin}^{\rm 2} (\theta)  +73\cdot  {\rm sin}^{\rm 3} (\theta)$  m s$^{-1}$. 

The latitude dependence of the meridional motion velocity can be better understood by considering sunspots of different Hale polarities. Figure \ref{fig3} shows the velocities for sunspots and pores of the leading polarity. For sunspots and pores, the latitude dependence is not monotonic.  In the low latitude region $\theta\approx 10 - 20^{\rm o}$  sunspots and pores move towards the equator. But at higher latitudes $\theta>20^{\rm o}$ the meridional displacement velocity is directed towards the poles. The velocities can be approximated by the pore dependencies: $v^{\rm pr}_{\rm ld}\approx 0.35-11.7\cdot {\rm sin}(\theta)+16.5\cdot {\rm sin}^{\rm 2} (\theta) +76.5\cdot {\rm sin}^3 (\theta)-32.7\cdot {\rm sin}^{4}(\theta )$ m s$^{-1}$;    for sunspots: $v^{\rm sp}_{\rm ld}\approx -0.35-18.3\cdot {\rm sin}(\theta)+32.2\cdot {\rm sin}^{\rm 2}(\theta) +71.4\cdot {\rm sin}^{\rm 3} (\theta)-6.7\cdot {\rm sin}^{\rm 4}(\theta)$ m s$^{-1}$.

Figure \ref{fig4} shows the dependence for sunspots and pores trailing polarity. Spots of trailing polarity have a relatively high velocity of shift to the poles at almost all latitudes. The dependence of the velocity of meridional movement of solar pores of trailing polarity on latitude can be described by a linear dependence on latitude $\theta$ in degrees: $v^{\rm pr}_{\rm tr}\approx 2.0+0.62\cdot \theta$  m s$^{-1}$. For sunspots, the approximation is as follows: $v^{\rm sp}_{\rm tr}\approx 0.02+0.94\cdot \theta$ m s$^{-1}$.

The speed of meridional movement of spots depends on their stage of development. We divided the stages of development into the growth stage and the decay stage. For this, we determined the time moment $t_{\rm max}$ of reaching the maximum area $S_{\rm max}$ by the spot. Then the growth stage will be for time intervals $t< t_{\rm max}$, and the decay stage $t> t_{\rm max}$. Figure 5a shows the dependences of the meridional displacement speeds at the area growth stage (growth) for sunspots of the leading and trailing polarity. For sunspots, the leading polarity dependences can be represented as: $v^{\rm gr}_{\rm ld}\approx - 2.3-26.7\cdot {\rm sin}(\theta)+40.9\cdot {\rm sin}^{\rm 2} (\theta) +91.4\cdot {\rm sin}^{\rm 3} (\theta)$ m s$^{-1}$. For sunspots with trailing polarity: $v^{\rm gr}_{\rm tr}\approx 3.5+18.5\cdot {\rm sin}(\theta)-38.5\cdot {\rm sin}{\rm 2} (\theta) +190.4\cdot {\rm sin}^{\rm 3} (\theta)$ m s$^{-1}$. For the pores of the leading polarity the dependence is as follows: $v^{\rm gr}_{\rm ld}\approx- 2.16-16.8\cdot {\rm sin}(\theta)+36.0\cdot {\rm sin}^{\rm 2} (\theta) +42.9\cdot {\rm sin}^{\rm 3} (\theta)$ m s$^{-1}$; for the trails, the pores are as follows: $v^{\rm gr}_{\rm tr}\approx 2.9+16.5\cdot {\rm sin}(\theta)-6.7\cdot {\rm sin}^{\rm 2} (\theta) +78.4\cdot {\rm sin}^{\rm 3} (\theta)$ m s$^{-1}$.

The meridional velocities for sunspots in the decay stage are shown in Figure 5b. The dependence of the velocity at the decay stage for sunspots: $v^{\rm dc}_{\rm ld}\approx 0.9-10.1 {\rm sin}(\theta)+10.4\cdot {\rm sin}^{\rm 2} (\theta) +58.8\cdot {\rm sin}^{\rm 3} (\theta)$ m s$^{-1}$;  $v^{\rm dc}_{\rm tr}\approx 1.4+3.5\cdot {\rm sin}(\theta)-22.6\cdot {\rm sin}^{\rm 2} (\theta) +85.3\cdot {\rm sin}^{\rm 3} (\theta)$ m s$^{-1}$.  For solar pores : $v^{\rm dc}_{\rm ld}\approx 0.31-7.9\cdot {\rm sin}(\theta)+17.5\cdot {\rm sin}^{\rm 2} (\theta) +23.2\cdot {\rm sin}^{\rm 3} (\theta)$ m s$^{-1}$;  $v^{\rm dc}_{\rm tr}\approx 0.12+6.4\cdot {\rm sin}(\theta)+19.9\cdot {\rm sin}^{\rm 2} (\theta) +83.9\cdot  {\rm sin}^{\rm 3} (\theta)$ m s$^{-1}$.  Comparing Figures 5a and 5b we can conclude that the meridional displacement rate at the growth stage has more pronounced latitude dependences.

By comparing Figures 5a and 5b we can conclude that the speed of meridional movement at the stage of spot area growth is more orderly.

\begin{figure}
               \includegraphics[width=0.48\textwidth,clip=]{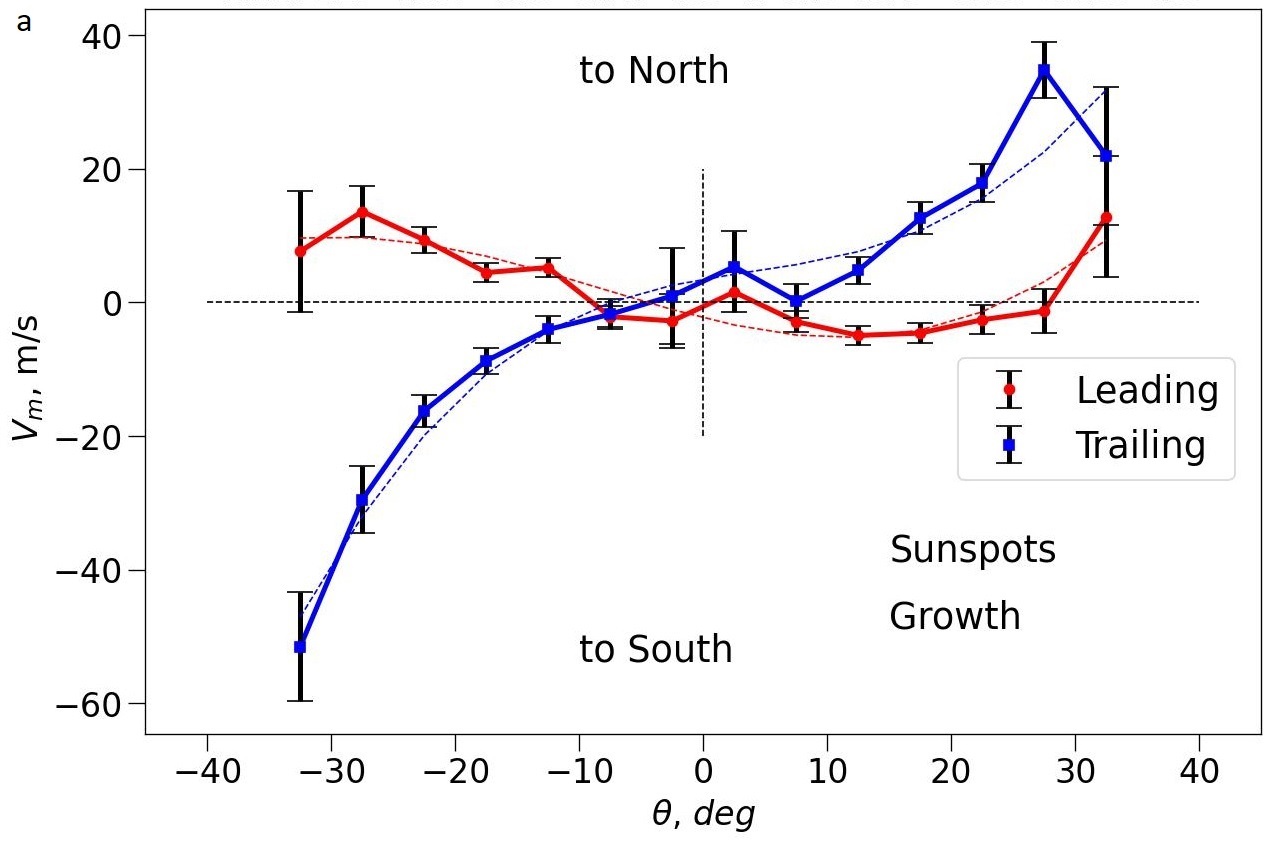}
               \hspace*{0.013\textwidth}
               \label{fig5a}
               \includegraphics[width=0.48\textwidth,clip=]{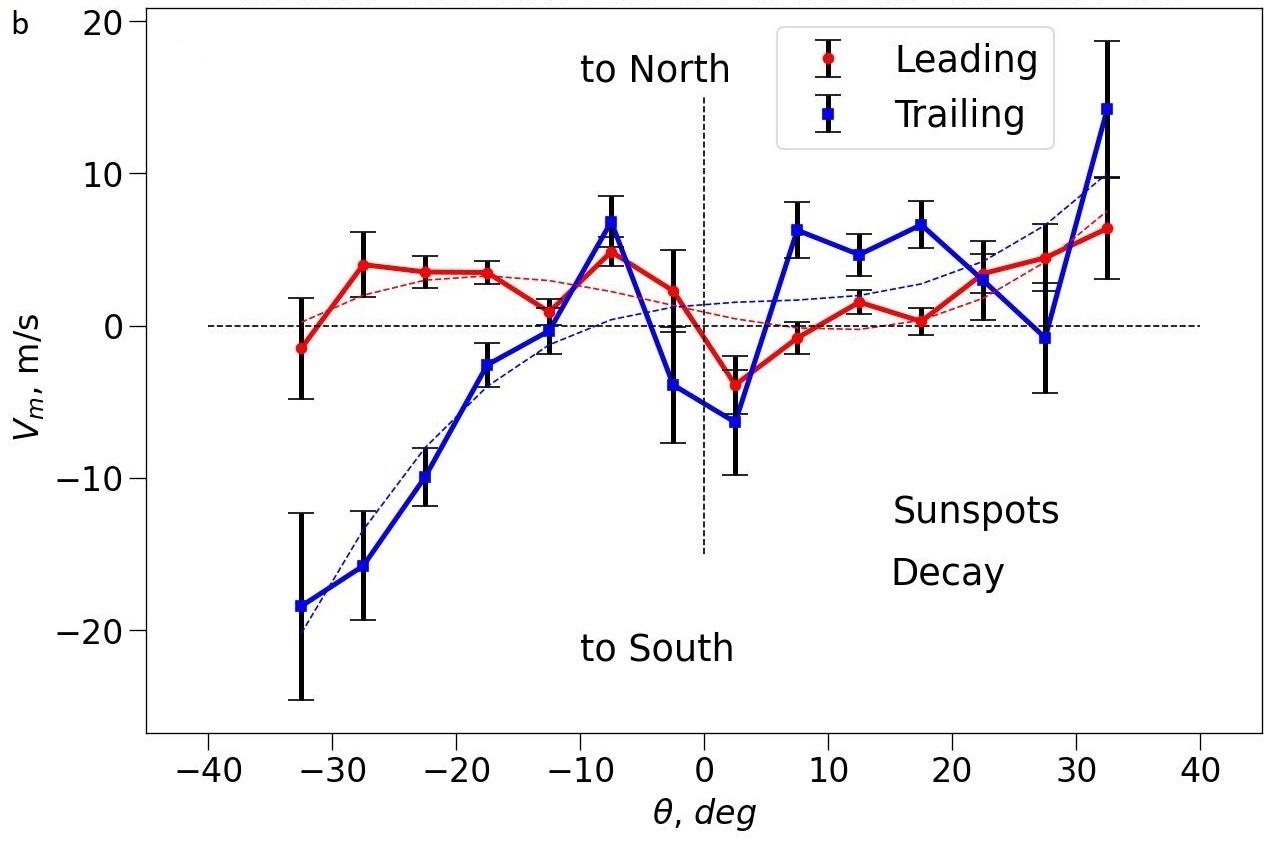}
               \label{fig5b}
\caption{The rate of meridional displacement of sunspots of leading  and trailing  polarity at the stage of area growth ({\it a, left}) and of sunspots of leading and trailing polarity at the decay ({\it b, right}) stage.   
}
\label{fig5}
\end{figure}

\begin{figure}
\centerline{\includegraphics[width=1.0\textwidth,clip=]{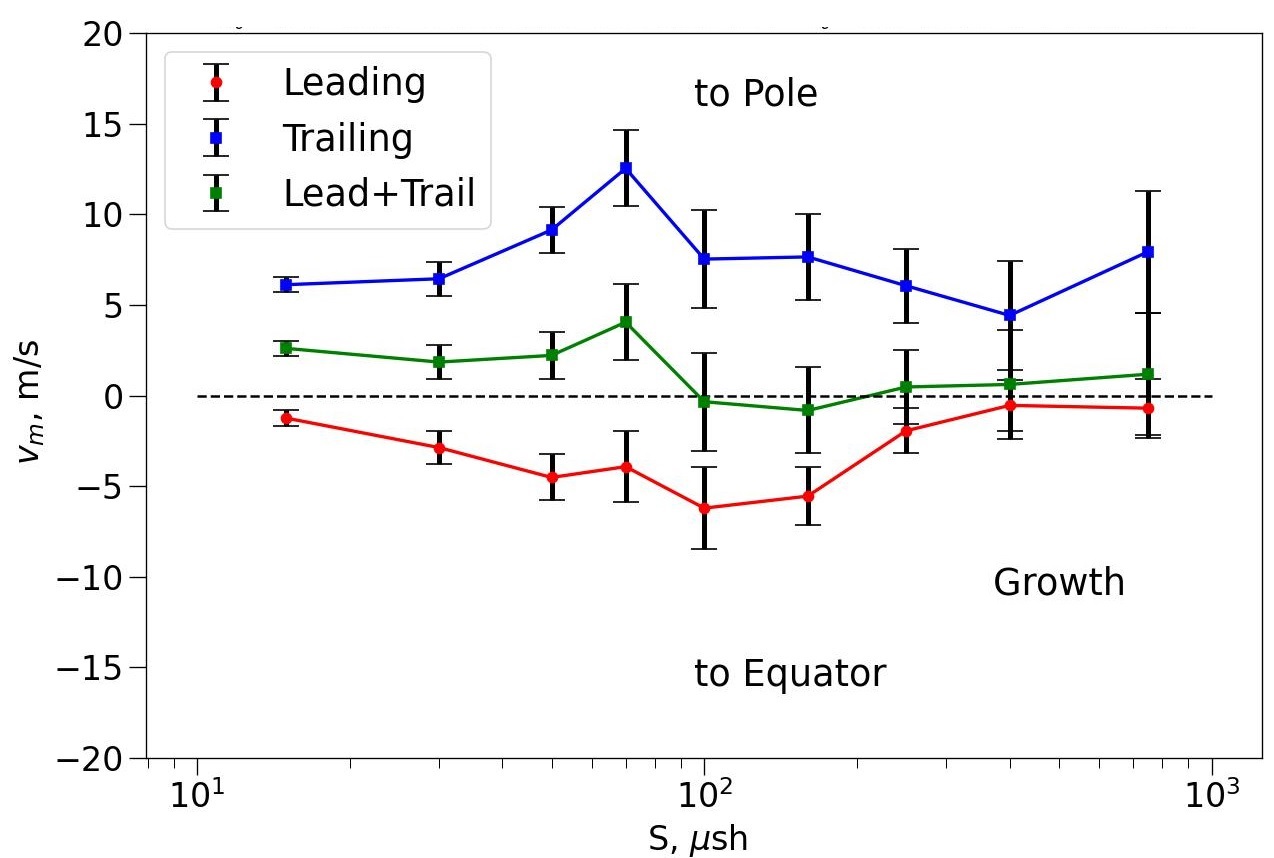} } 
\caption{Speeds of meridional motion of sunspots, depending on their area. All spots ({\it green}) and spots of leading ({\it red}) and trailing ({\it blue}) polarity are shown separately.}
\label{fig6}
\end{figure}

Consider changes in meridional velocities depending on the area of sunspots. Figure \ref{fig6} shows the velocities of the meridional displacement $v_{\rm m}$ of sunspots for different ranges of areas $S$ For this sample, we used the entire range of latitudes of sunspot existence during the sunspot growth stage.  Figure \ref{fig6} shows the values of the velocities for all the spots and separately for the spots of the leading and trailing polarity of the magnetic field. For clarity, we took the velocities of spots in the southern hemisphere with the opposite sign. This allowed us to distinguish the movements directed to the poles and the movements directed to the equator.  The averaged velocity over all areas of the spots of the trailing polarity  was $v_{\rm tr} \approx 6$ m s$^{-1}$, while that of the spots of the leading polarity was $v_{\rm ld}\approx -3$ m s$^{-1}$. This indicates different directions of velocities for spots of leading and trailing polarity. Spots of trailing polarity move predominantly to the poles, and spots of leading polarity shift predominantly to the equator. Up to area values of $S\approx 100$ $\mu$sh, the absolute values of velocities increase, and then they begin to decrease. The confidence intervals are also presented in Figure \ref{fig6}.

Linear approximations of the latitude dependence of the sunspot velocity at the growth stage for spots of leading and trailing polarity depending on the area are presented in Figures 7a, 8a. Here, averaging was performed in 10-degree latitude intervals. For all considered area ranges, the approximations for spots of trailing polarity have a positive angle, i.e., they are mainly directed toward the poles.  Velocity approximations for spots of leading polarity have a negative angle, i.e. they are mainly directed toward the equator. Figures 7b and 8b show the coefficient values of the velocity approximation formulas for the leading and trailing polarity spots in the growth stage. Maximum absolute values are achieved for sunspots with an area of $S\approx 80\,-\,100$ $\mu$sh.

\begin{figure}
               \includegraphics[width=0.48\textwidth,clip=]{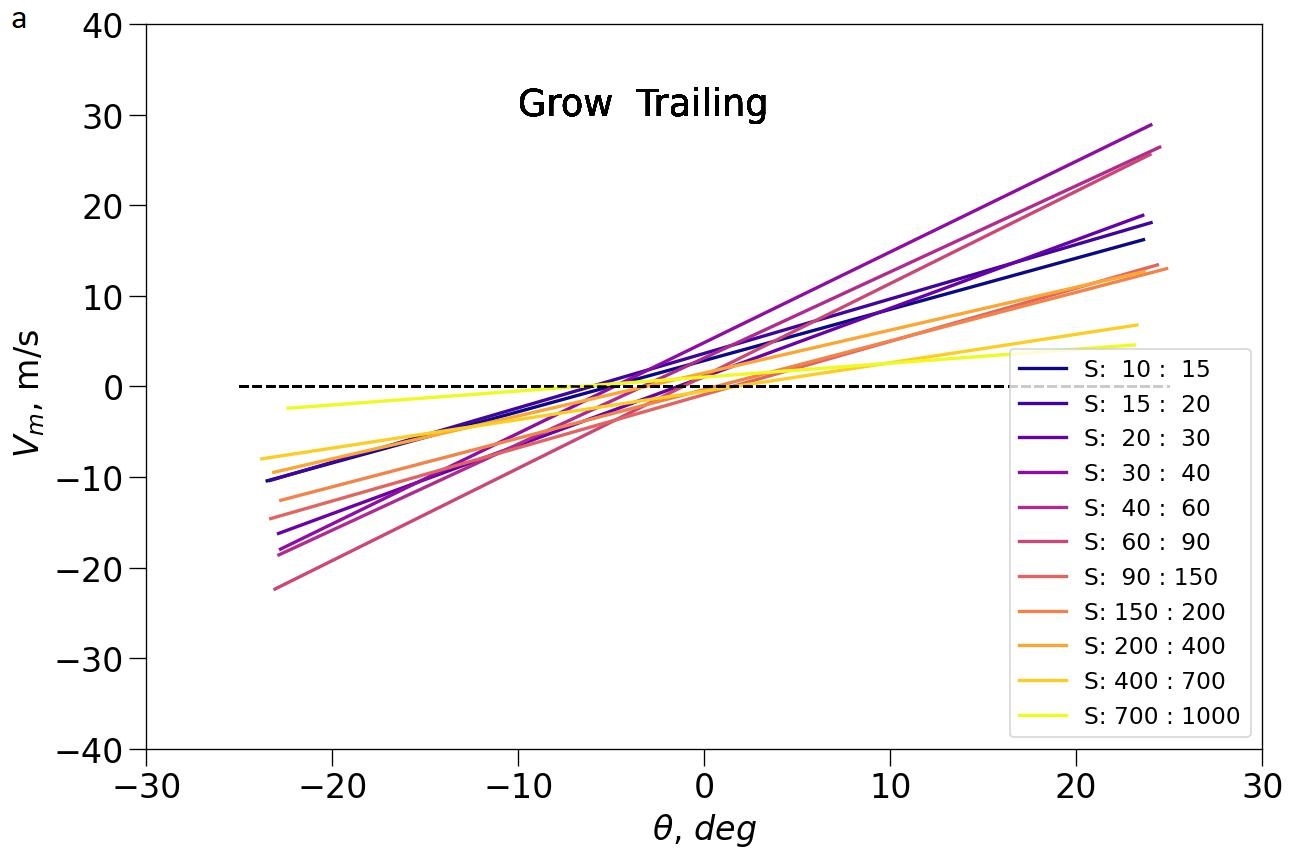}
               \hspace*{0.013\textwidth}
               \includegraphics[width=0.48\textwidth,clip=]{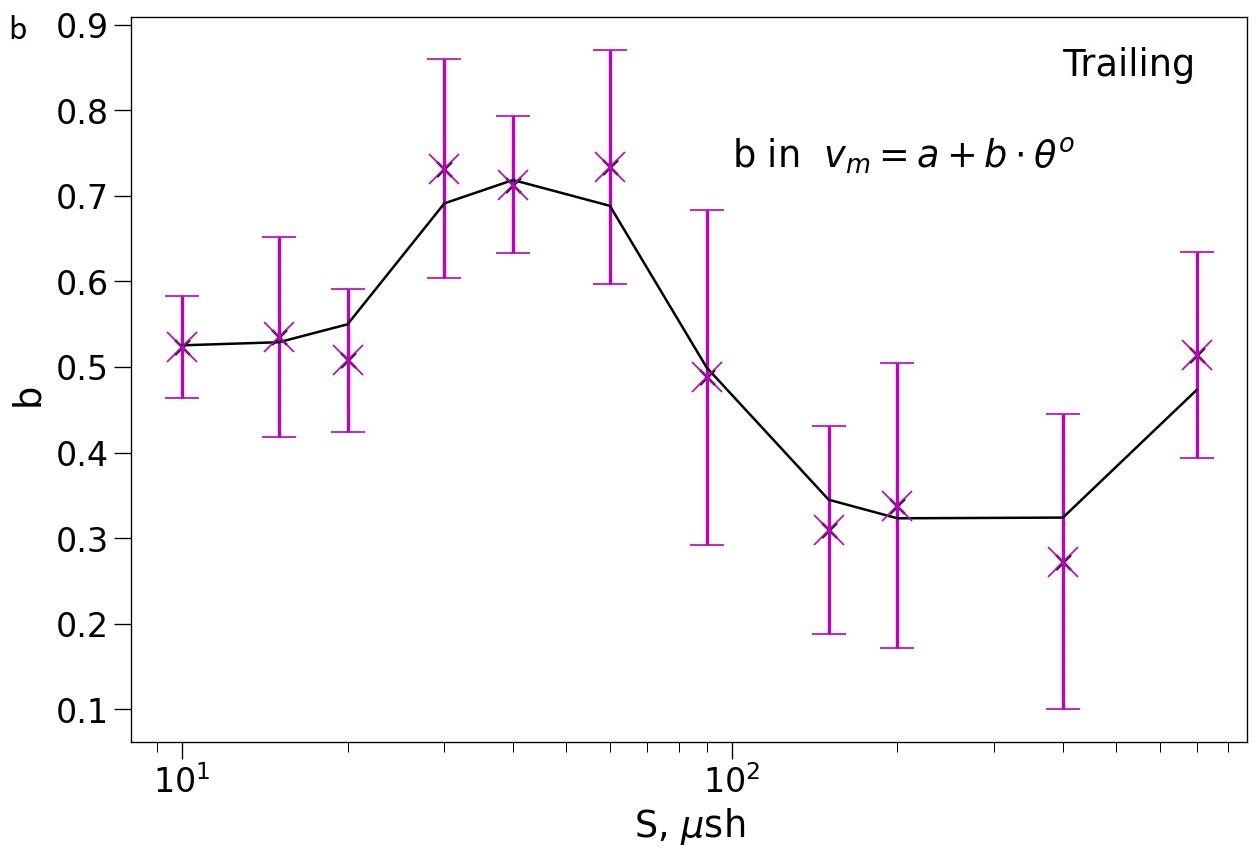}
\caption{ ({\it a, left}) Linear approximations for sunspots in the growth stage for different area intervals for sunspots of the trailing polarity. ({\it b, right}) The coefficient b in the linear approximation in the formula $v_{m}=a+b\cdot \theta ^{\rm o}$, m s$^{-1}$.
}
\label{fig7}
\end{figure}

\begin{figure}
               \includegraphics[width=0.48\textwidth,clip=]{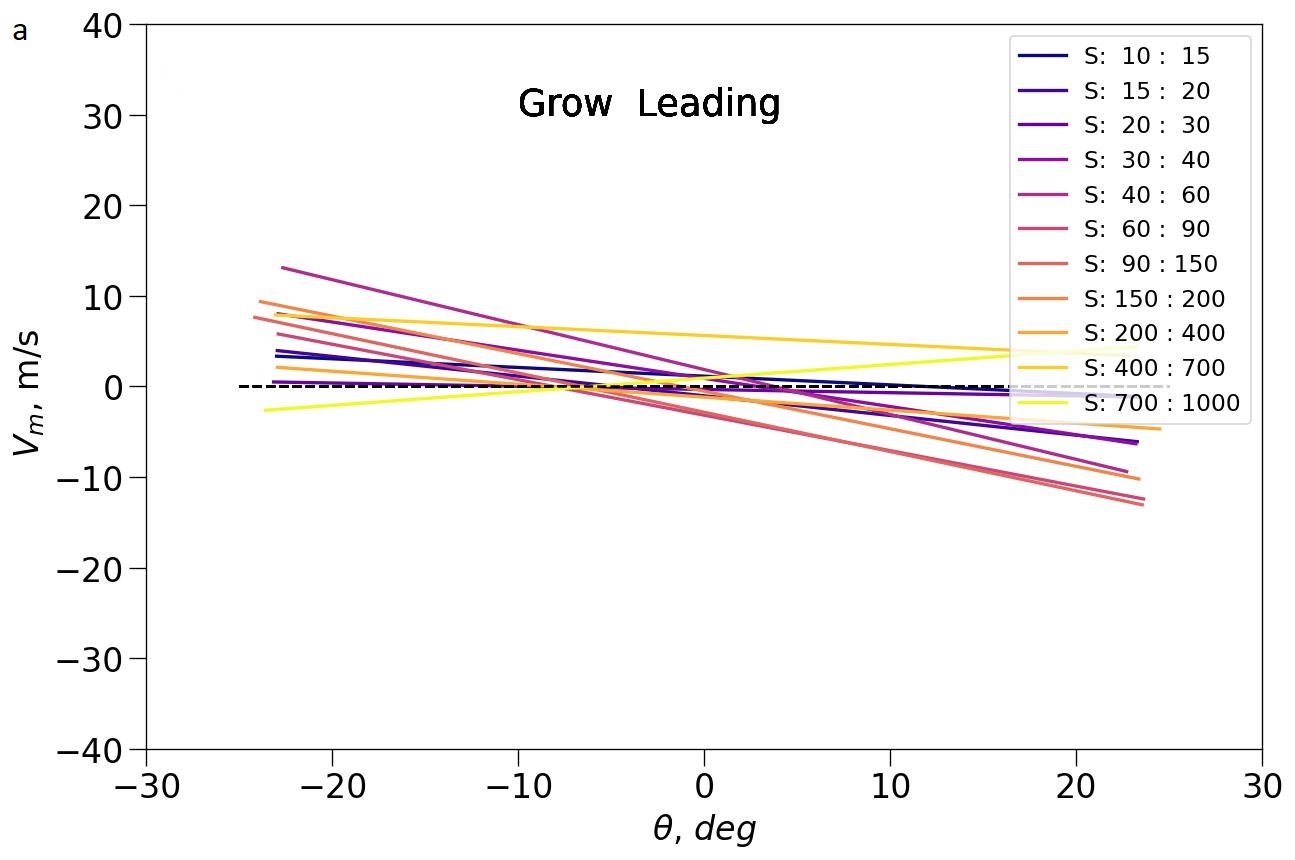}
               \hspace*{0.013\textwidth}
               \includegraphics[width=0.48\textwidth,clip=]{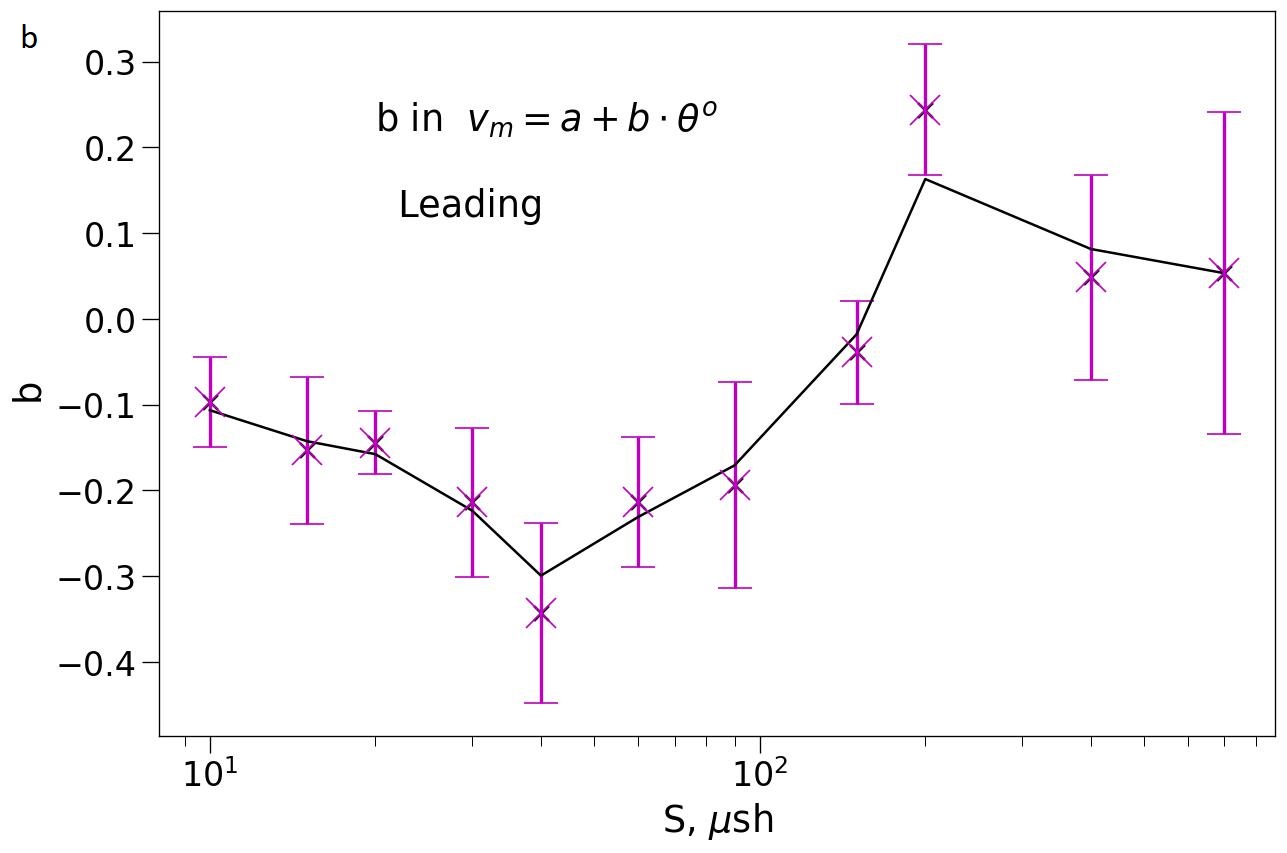}
\caption{Same as Figure ~\ref{fig7}, but for sunspots of leading polarity.}
\label{fig8}
\end{figure}

\section{Discussion}

In this paper, we analyze the velocities of individual sunspots and pores. We find dependences of the velocities on latitude. Pores and sunspots, if their polarity corresponds to the trailing polarity, on average move in the direction of the poles and show an increase in meridional velocity with increasing latitude (Figure \ref{fig3}).
The average velocity values obtained reach values of 20\,-\,40 m s$^{-1}$ at latitudes of $\theta\approx 30^{\rm o}$. This is slightly less than the values obtained by other authors \citep{Schunker, Weber}, but we used a much larger number of measurements, which ensured that more obvious dependences on latitude were obtained.

The velocity of the meridional movements is significantly dependent on the Hale polarity of the magnetic field (Figures \ref{fig3}, \ref{fig4}). This difference is associated with proper movements of the emerging flux tube \citep{Schunker,TT24}. 
Thus, in \cite{TT24} it was found that the rate of differential rotation of the leading polarity spots is higher than the rate of rotation of the trailing polarity spots. The difference can be explained by the emergence of an expanding magnetic field tube (\citeauthor{TT24} \citeyear{TT24},  Figure 8). Since the active regions (AR) are inclined relative to the equator, the divergence of spots of different polarity should lead not only to a difference in rotation rate but also in the velocity of meridional movements. The magnitude of the velocity should vary depending on the angle of inclination of the AR, which in turn depends on the latitude of the active regions, according to Joy's law. We see this in Figures \ref{fig3}, \ref{fig4}. Another explanation could be different meridional velocities of matter flow in the convective zone, which differently affect the velocities of movement of the leading and trailing polarities. These hypotheses can be tested by studying the relationship between meridional displacements and the AR tilt angle. We plan to do this in the near future. In any case, different meridional displacement velocities allow us to better understand the poleward drift of predominantly trailing polarity magnetic fields, which is important for modeling the solar dynamo. 
The proper velocities of the spots of leading and trailing polarity are added to the velocity of the global meridional circulation. For spots of trailing polarity, this leads to an increase in velocity in the direction toward the poles (Figure  \ref{fig3}). For spots of leading polarity, these velocities are in different directions.  At low latitudes velocities are directed toward the equator. At high latitudes, the global circulation velocity exceeds the divergence velocity when the force tube emerges, and the velocity becomes directed toward the poles (Figure  \ref{fig4}).


From Figure \ref{fig6} and Figures  \ref{fig7}, \ref{fig8}  it follows that the velocities of the meridional movements are highest for the area interval $S\approx 80\,-\,100$ $\mu$sh. This is typical for both sunspots of trailing polarity and sunspots of leading polarity. Perhaps this is due to the fact that the bases of the emerging magnetic field tube have the highest velocities for this area range.

\begin{fundinginformation}
We acknowledge the financial support of the Ministry of Science and Higher Education of the Russian Federation, grant number 075-03-2025-420/4.

\end{fundinginformation}

\begin{acknowledgments}
Data are courtesy of NASA/SDO and the HMI science team.
\end{acknowledgments}

\begin{dataavailability}
All data used are available at the HMI JSOC.

\end{dataavailability}

\bibliography{spot_moution.bib}{}
\bibliographystyle{aasjournal}

\end{document}